\documentclass[11pt]{article}

\usepackage{graphicx}
\usepackage{epstopdf}
\usepackage{amsmath}
\usepackage{amssymb}
\usepackage{amsfonts}
\usepackage{amsthm}
\usepackage[usenames]{color}
\usepackage{array}
\usepackage{mathrsfs}
\usepackage{epic,eepic,pstricks}
\usepackage{rotating}
\newcommand{\scri}{{\mathscr I}}

\usepackage[
      colorlinks=true,
      linkcolor=blue,
      urlcolor=blue,
      filecolor=blue,
      citecolor=red,
      pdfstartview=FitV,
      pdftitle={},
      pdfauthor={},
      pdfsubject={},
      pdfkeywords={},
      pdfpagemode=None,
      bookmarksopen=true
]{hyperref}

\usepackage{epsfig}

\usepackage{hyperref}

\newcommand{\bea}{\begin{eqnarray}}
\newcommand{\eea}{\end{eqnarray}}
\newcommand{\ba}{\begin{array}}
\newcommand{\ea}{\end{array}}
\newcommand{\ee}{\end{equation}}

\numberwithin{equation}{section}

\begin{document}

\begin{flushright}
\texttt{\today}
\end{flushright}

\begin{centering}

\vspace{2cm}

\textbf{\Large{ Complexity Growth in Flat Spacetimes }}

  \vspace{0.8cm}

  {\large Reza Fareghbal, Pedram Karimi }

  \vspace{0.5cm}

\begin{minipage}{.9\textwidth}\small
\begin{center}

{\it  Department of Physics, 
Shahid Beheshti University, 
G.C., Evin, Tehran 19839, Iran.  }\\

  \vspace{0.5cm}
{\tt  r$\_$fareghbal@sbu.ac.ir, pedramkarimie@gmail.com}

\end{center}
\end{minipage}


\begin{abstract}
 We use the complexity equals action  proposal to calculate the rate of complexity growth for field theories that are the holographic duals of asymptotically flat spacetimes.  To this aim, we evaluate the on-shell action of asymptotically flat spacetime on the Wheeler-DeWitt patch.  This results in the same expression as can be found by taking the flat-space limit from the corresponding formula related to the asymptotically AdS spacetimes. For the bulk dimensions that are greater than three, the rate of complexity growth at late times approaches from above to  Lloyd's bound. However, for the three-dimensional bulks, this rate is a constant and differs from  Lloyd's bound by a logarithmic term.  
\end{abstract}

\end{centering}

\newpage



\section{Introduction}
It was proposed in \cite{Bagchi:2010zz,Bagchi:2012cy} that the holographic dual of asymptotically flat spacetimes in $d+1$ dimensions is a $d$-dimensional field theory that has BMS symmetry. These field theories are known as BMSFT. From the point of view of the bulk theory,  BMS symmetry is the asymptotic symmetry of the asymptotically flat spacetimes \cite{BMS}-\cite{aspects}. In three and four dimensions these symmetries are infinite dimensional. In the one-dimension lower boundary theory, this symmetry is given by contraction of conformal symmetry. In this view, one can study flat-space holography by starting from AdS/CFT and takeing the appropriate limit. The flat-space limit of the bulk theory corresponds to the ultrarelativistic limit of the boundary CFT \cite{Bagchi:2012cy}.  

Since BMS symmetry is infinite dimensional, it is possible to find the universal  aspects of BMSFTs that are  independent of the action and details of the theory. In \cite{Bagchi:2012xr}, a Cardy-like formula has been proposed for BMSFT$_2$. This formula gives an estimation for the degeneracy  of the states of this field theory. The interesting point is that this formula yields the entropy of three-dimensional flat space cosmology (FSC), which is given by taking the flat space limit from the BTZ black holes. The universal structure of the correlation functions of BMSFT$_2$ and BMSFT$_3$ has been studied in \cite{Bagchi:2015wna}-\cite{Fareghbal:2018xii}. The entanglement entropy formula and also the holographic interpretation of this formula in the context of flat/BMSFT have been studied in \cite{Bagchi:2014iea}-\cite{Hijano:2018nhq} . In all of the above mentioned works, the calculations that are  done in asymptotically flat spacetimes nicely fit to the results given by taking the ultrarelativistic limit of CFTs. For an almost complete list of   papers related to the flat-space holography  see the references of \cite{Riegler:2017fqv} and \cite{Prohazka:2017lqb}. 

  After the remarkable work of Ryu and Takayanagi \cite{Ryu:2006bv} (which proposes a holographic description for the entanglement entropy of CFT in the context of AdS/CFT), it seems that we can translate all of the information physics to the gravitational counterpart by the virtue of holography. There are other aspects of information physics that seem natural to find their holographic picture. One of the most important physical quantities in information physics is complexity (see \cite{complexity 1,Aaronson:2016vto} for reviews). The complexity measures the number of gates that are needed to achieve a desirable state from an initial state. There are two different proposals for the holographic complexity. Here, we will focus on the complexity equals action (CA) conjecture  that was proposed in \cite{Brown:2015bva, Brown:2015lvg}. According to this proposal, the boundary complexity is given by the bulk gravitational action that is evaluated on a region of spacetime known as the Wheeler-DeWitt (WDW) path. It is a portion of space-time bounded by null surfaces anchored at the related time on the boundary.   There is a different proposal  that relates the complexity to the volume of an anchored region \cite{Stanford:2014jda, Alishahiha:2015rta, Barbon:2015ria, Barbon:2015soa} (complexity=volume (CV) proposal) . Both of these conjectures have been proposed in the context of AdS/CFT correspondence.
 
 In this paper, we want to use the CA conjecture and calculate the rate of complexity growth in BMSFT by using flat-space holography. As mentioned above, an approach for improving flat-space holography is given by taking the flat-space limit from the AdS/CFT calculations\footnote{ For the CV conjecture in $d>2$, the flat-space limit was already shown to work by Susskind (see \cite{Susskind:2014moa}). Consequently,  one may expect that the CA conjecture admits a regular flat-space limit. }. The corresponding computation of the rate of complexity growth in the context of AdS/CFT has been done in \cite{Carmi:2017jqz} (see also \cite{Cai:2016xho}-\cite{Moosa:2017yiz}). Therein, the gravitational action is evaluated in the background of eternal two-sided black holes. It was assumed that these geometries are the holographic duals of thermofield double states in the boundary theory \cite{Maldacena:2001kr}. 
 
   The final answer of \cite{Carmi:2017jqz} for the rate of complexity growth is given in terms of bulk parameters. It is not difficult to take the flat-space limit from these results. One can check that taking the flat-space limit from the results of \cite{Carmi:2017jqz} yields the following expressions for the  rate of complexity growth: \footnote{ For obtaining \eqref{result d3} we have assumed that $R=L$ where $L$ is the AdS radius and $R$ is the radius of periodic coordinate in the boundary geometry. Without this choice for the parameter $R$, the flat space limit of (2.60) in \cite{Carmi:2017jqz} is not well-defined. }
 
 \begin{equation}\label{result for d4}
\dot{\mathcal{C}} = \frac{1}{\pi}\frac{d I}{d \mu}=\frac{1}{\pi}\left[ 2 M - \frac{r_M^{d-2} \Omega_{d-1} (d-1)}{16 \pi G_{N}  }f(r_M)\log \frac{-\mathfrak{a}^2}{f(r_M)}\right], \qquad d\geq 3,
\end{equation} 

 \begin{equation}\label{result d3}
\dot{\mathcal{C}}= \frac{1}{\pi} \left( 2 M+ M \log \frac{\mathfrak{a}^2}{8 G_{N} M}\right), \qquad d=2.
\end{equation}
 
 The parameters appearing in these formulas are explained in section 3. According to \,\,\,\, flat/BMSFT correspondence, Eq. \eqref{result for d4} is the rate of complexity growth for BMSFT$_{d}$, $d \geq 3$, and Eq. \eqref{result d3} is the  same rate for BMSFT$_2$. Our goal in this paper is to directly calculate both of these formulas by using the CA proposal in asymptotically flat spacetimes. The background geometries that we use in this paper are asymptotically flat two-sided black holes in spacetime dimensions greater than three and two-sided FSC in three dimensions. All of these geometries are given by taking the flat-space limit from their corresponding asymptotically AdS counterparts.  The on-shell action in the flat case is evaluated on a particular region of spacetime, which is given by the intersection of two WDW patches.  The null surfaces bounding these patches are anchored on the future or past null infinity. However, their intersection points meet  neither past nor future null infinity.    We show that despite the vanishing bulk term in the on-shell action, the results \eqref{result for d4} and \eqref{result  d3} are deducible from the boundary and joint terms.

The paper is organized as follows: In section two we start from preliminaries.  Section three and four  include the main part of our calculations, and we directly evaluate the rate of complexity growth in BMSFTs by using flat space holography in, respectively, $d\geq 3$ and $d=2$ dimensions. The last section is devoted to discussions.

\section{Preliminaries} 

In this section we use the flat-space holography to compute the rate of  complexity growth of BMSFT. We use the CA proposal for BMSFT$_2$ and BMSFT$_3$, which requires computation of the on-shell action for, respectively, three- and four- dimensional asymptotically flat geometries. In this paper we consider static solutions with line element
\begin{equation}
ds^2 = - f(r) dt^2+\frac{dr^2}{f(r)} +r^2 d \Omega^2_{d-1},
\end{equation}
where  $f(r)=-8 G_N M$ for $d=2$ and for $d\geq 3$ it is given by 
\begin{equation}
\label{schwmet}
f(r) = 1 - \frac{\omega^{d-2}}{r^{d-2}}, \qquad M= \frac{d-1}{16 \pi G_{N}} \Omega_{d-1} \omega^{d-2},
\end{equation}
where M is the mass parameter and $\Omega_{d-1}$ is the volume of a $(d-1)$-dimensional unit sphere.  It will prove convenient  to use $(u,r^{\ast})$ or $(v,r^{\ast})$ coordinates instead of $(t,r)$ where

\begin{equation}
\label{tort}
r^{\ast}(r) = \int \frac{dr}{f(r)},\qquad v= t + r^{\ast}, \qquad u= t-r^{\ast}.
\end{equation}
$v$ and $u$ are, respectively, the advanced and retarded times, and $r^{\ast}$ is the tortoise coordinate. 
 It is important to note here that at  $r=r_{h}$ where  $r_h$ is the root of $f(r)$, $r^{\ast}$ gets its minimum value,
\begin{equation}\label{rmin}
r^{\ast}(r_h) = r_{min} \simeq -\infty.
\end{equation}

According to the proposal of \cite{Brown:2015lvg}, the complexity of dual theory is given by the gravitational action  evaluated on a region of spacetime known as the WDW patch. The WDW patch is given by the union of all the spatial slices anchored at a given boundary time \cite{Susskind:2014rva}. Here we use this definition and impose it in the flat space. In the flat space, as it was shown in figure 1, the  WDW patch is  the intersection of spatial slices anchored at future or null infinity. It is clear that the WDW patch in the flat scenario connects to the infinity via the null geodesics and does not reach it. A similar situation   happens in the holographic  description of the BMSFT entanglement entropy  where the minimal surface does not reach the boundary and connects to it  via two null geodesics \cite{Jiang:2017ecm}. Thus our prescription is  a natural extension of the WDW patch definition and  also is consistent with the  holographic description of the entanglement entropy in flat spacetimes.

   In this paper we consider asymptotically flat geometries which are given by taking the flat space  limit from the asymptotically AdS eternal two-sided black holes. Thus we have right and left null infinities in the Penrose diagram of these spacetimes. In order to control  divergent terms we need to restrict the WDW patch by using some cutoffs. Therefore the boundary of space-time on which the on-shell action must be computed consists of null surfaces besides timelike ones and their joint points. A complete computation requires  that we accompany boundary terms to the bulk action. Hence we use the following generic action:
\begin{align}
\nonumber
I =& \frac{1}{16 \pi G_{N}} \int_{\mathcal{M}} d^{d+1} x \sqrt{-g}~\mathcal{R}
\\\nonumber
 &+ \frac{1}{8 \pi G_{N}} \int_{\mathcal{B}} d x^d  \sqrt{|h|} ~K + \frac{1}{8 \pi G_{N}} \int _{\Sigma}d^{d-1}x \sqrt{\sigma} ~\eta
 \\\label{action}
 &+\frac{1}{8 \pi G_{N}} \int_{\mathcal{B'}} d x^d \sqrt{\gamma} ~\kappa + \frac{1}{8 \pi G_{N}} \int _{\Sigma}d^{d-1}x \sqrt{\sigma} ~a
\end{align}
 The first term is related to the volume of the WDW patch and is vanishing in the flat scenario. The Vanishing of the bulk term in the on-shell action is the most important technical difference between the AdS case and the flat case.

The second line of action belongs to the non-null boundaries. The first term is known as the Gibbons-Hawking-York (GHY) term in the spacelike and timelike sector of the boundary. The GHY term guarantees a well-defined variation principle with the Dirichlet boundary term. The second term belongs to the joint term that is evaluated at the intersection of two non-null hypersurfaces.

In the third line, we  encounter null hypersurfaces. The null boundary term  gained some attention recently. The first term is the counterpart of the GHY term in the null boundary. This term can always be ignored by assuming an affine null parameter. The second term evaluates joint terms in the intersection of two hypersurfaces where at least one of the hypersurfaces is null.

We use the instruction of \cite{Carmi:2016wjl} to evaluate  terms of \eqref{action}. The boundary terms for null hypersurfaces were discussed in several works \cite{Parattu:2015gga,Lehner:2016vdi}. The joint terms first introduced by Hayward \cite{Hayward:1993my} for spacelike and timelike boundaries were extended by \cite{Lehner:2016vdi} to the null hypersurfaces. It is notable that neither  the boundary terms nor the joint terms depends on the cosmological constant. It is worth mentioning that in the context of  holographic renormalization  the counterterms that cancel the divergent terms in the action are related extremely to the existence of the cosmological constant\cite{Skenderis:2002wp, Costa:2013vza}. As \cite{Costa:2013vza} observed, local counterterms in asymptotically AdS spacetimes become nonlocal in the asymptotically flat spacetimes. To our knowledge, the holographic renormalization of asymptotically flat spacetimes is still an open problem.

Using \eqref{schwmet} we can calculate the terms of \eqref{action}. The null boundary term vanishes because we  can always choose a null parameter to be affine, and then the null boundary terms  in  \eqref{action} do not contribute to the on-shell action. It remains the GHY term that has its contribution from the timelike or the spacelike surfaces
\begin{align}
\label{GHYspace}
I_{GHY}^{spacelike} &= -\int dt~ \frac{r^{d-1}~ \Omega_{d-1}}{16 \pi G_{N}} \left( f'(r) +\frac{2(d-1)}{r} f(r) \right)\Bigg|_{\text{r=const}},
\\
\label{GHYtime}
I_{GHY}^{timelike} &= \int dt~ \frac{r^{d-1} ~\Omega_{d-1}}{16 \pi G_{N}} \left( f'(r) +\frac{2(d-1)}{r} f(r) \right)\Bigg|_{\text{r=const}}.
\end{align}
In our calculation in the rest of this paper, all of the joint terms have at least  one null part. Hence,  it is adequate to compute the last term of \eqref{action},
\begin{align}
\label{jointspace}
I_{J}^{spacelike} &= \frac{\Omega_{d-1} ~r^{d-1}}{16 \pi G_{N}} \log |f(r)|,
\\
\label{jointtime}
I_{J}^{timelike} &= \frac{-\Omega_{d-1}~ r^{d-1}}{16 \pi G_{N}} \log |f(r)|,
\\
\label{jointnull}
I_{J}^{null} &= \frac{-\Omega_{d-1} ~r^{d-1}}{8 \pi G_{N}} ~\text{Sign}(f(r)) ~\log \frac{\mathfrak{a}^2}{|f(r)|},
\end{align}
where  all of the joint points are labeled by their second non-null leg. The null joint term has an ambiguity  due to the normalization constant of the null vectors $\mathfrak{a}$. This ambiguity is the same as the ambiguity in the AdS case \cite{Carmi:2017jqz} and reveals the existence of the new length in BMSFT. 


\section{BMSFT Complexity Gowth in $d\geq3$  }
\subsection{Initial time}
The Penrose-Cartan diagram of the asymptotically flat two sided black hole is depicted in figure 1. The region of spacetime on which gravitational action is evaluated is shown by the gray color.
We  impose some cutoffs to the problem. The first type is a cutoff surface at $r^{\ast}=\epsilon_{0}$, which takes place near two singularities. The second type of cutoffs  mentioned as UV cutoffs take place at $r^{\ast}= r_\text{max}$ near the position of the dual field theories. It is clear  from the Penrose-Cartan diagram that the intersection of WDW patches (depicted by gray) never meets UV cutoffs. This is another difference between the computation of the complexity growth in the AdS holography and the flat-space  holography.

The geometry of the Penrose diagram reveals that the boundary times on the left-  and right-hand sides are minus   each other. 
We denote the times of relevant points in the null infinities as
\begin{align}
\scri^{+} \Rightarrow\begin{cases}
                                             \beta=  u^{+}_{R},
                                              \\
                                               \alpha= v^{+}_{L} 
                                             \end{cases} \qquad \scri^{-} \Rightarrow\begin{cases}
                                             \lambda= v^{-}_{R},
                                              \\
                                              \sigma= u^{-}_{L}.
                                             \end{cases}
\end{align}
 From now on the  indices   $\pm$ refer, respectively,  to   $\scri^+$ and $\scri^-$.

In order to compute the complexity growth, we need to consider  evolution of the gray region. Since we want to compare our results with \eqref{result for d4} which is given by taking the flat-space limit from the AdS case, we  assume that the BMSFT on the right- and left-hand sides develop symmetrically. This requires a symmetric evolution of the advanced and retarded coordinates on the different null infinities as
\begin{equation}
 u^{+}_{R}= - v^{+}_{L} =\mu^+,
\qquad  v^{-}_{R}= - u^{-}_{L} = \mu^-.
\end{equation}

 There is a critical time when the gray region leaves the cutoff near the past singularity. For the simplicity of calculation and avoiding unnecessary shifts in the origin of boundary times, we assume that this cross occurs at $t=0$. The symmetric evolution guarantees that the last crossing point remains permanently on $t=0$.    The initial times are those before this time. It is clear from the Penrose diagram that for the symmetric evolution the initial and late  times are, respectively, given  by $\mu^{\pm} < \mu^{\pm}_{c}$ and $\mu^{\pm} > \mu^{\pm}_{c}$ where 
\begin{equation}\label{criticaltime}
\mu^+_{c} =- r^{\ast}(\epsilon_{0}).
\end{equation}

 All of the relevant points in the Penrose diagram are collected in the next table:\\
 
\begin{center}
\begin{tabular}{ |p{3cm}|p{3cm}|p{3cm}|p{2cm}|p{3.7cm}|  }
 \hline
 
 \multicolumn{4}{|c|}{The points on Penrose-Cartan diagram in figure 1} \\
 \hline
 {Point's ame} & \qquad \quad{t} &\qquad\quad {${r ^{\ast}}$ }& {{Sign of $f(r)$}}\qquad  \\
 \hline
\qquad\quad X &\quad $r^{\ast}(\epsilon_{0}) - \mu^-$ & \qquad$r^{\ast}(\epsilon_{0})$  &\qquad - \\
 \hline
\qquad\quad W &\quad $-r^{\ast}(\epsilon_{0}) +\mu^-$ &\qquad $r^{\ast}(\epsilon_{0})$  &\qquad - \\
 \hline
\qquad\quad Y &\quad $r^{\ast}(\epsilon_{0}) + \mu^+$ & \qquad$r^{\ast}(\epsilon_{0})$  &\qquad -  \\
 \hline
\qquad\quad Z &\quad $-r^{\ast}(\epsilon_{0}) - \mu^+$ &\qquad $r^{\ast}(\epsilon_{0})$  &\qquad -  \\
 \hline
\qquad\quad P &\quad $\frac{-1}{2}( \mu^+ + \mu^-)$ &\quad $\frac{1}{2}( \mu^- - \mu^+)$ &\qquad + \\
 \hline
\qquad\quad Q &\quad  $\frac{1}{2}(\mu^+ + \mu^-)$ &\quad $\frac{1}{2}( \mu^- - \mu^{+})$  &\qquad + \\
 \hline
 \end{tabular} 
 \end{center}
\vspace{.5cm}
In the calculation  of  \cite{Carmi:2017jqz} for the asymptotically AdS black holes the  complexity growth is evaluated in the time that   is given by adding  left and right times. For the asymptotically flat cases, besides left and right development, the lower and upper sides of the WDW path can develop independently. The origin of this difference is that the times of past and future null infinities are given by advanced and retarded times. In order to reproduce the results that are given by taking the flat space limit, we  have to consider symmetric evolution on the future and past null infinities. Precisely, we need to calculate the rate of complexity growth with respect to $\mu$ where     
 \begin{equation} \label{symmetricquantity}
  \mu^+ = \mu^- + \chi,  \qquad \mu^+ +\mu^- =\mu,
\end{equation}
where $\chi$ is a constant. In  Appendix A we calculate the nonsymmetric evolution by considering the $\mu^+ = \gamma \mu^-+\chi$ case. The results of taking the flat space limit are given when $\gamma=1$.

At this point, we have all of the requirements  to evaluate on-shell action \eqref{action} for the gray part of figure 1. There are six different GHY terms and four different joint terms:
\begin{figure}
\label{fig:fig1}
\begin{center}
\includegraphics[scale=0.4]{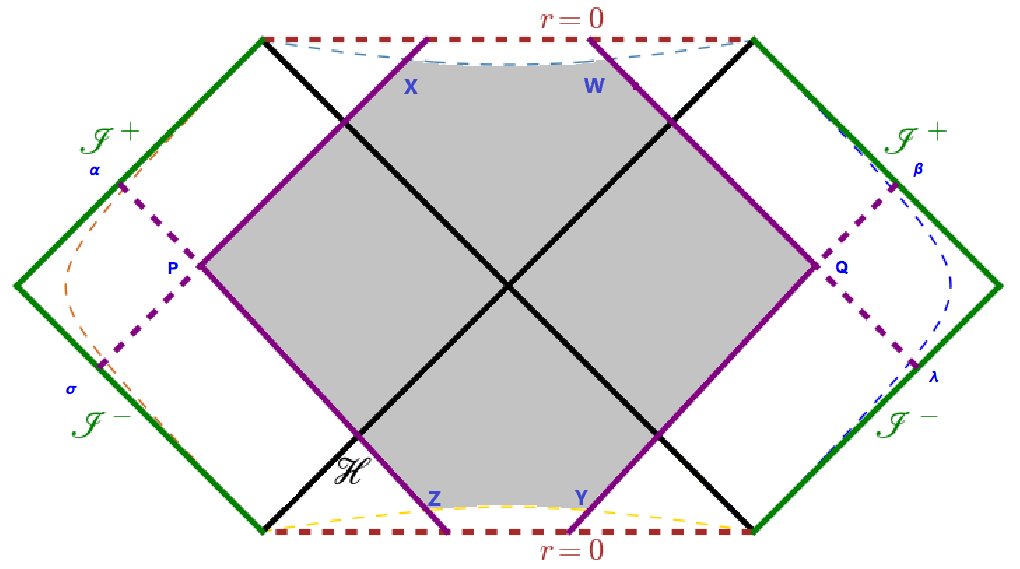}
\caption{ Initial time for $d>3$. Gravitational action is evaluated on the gray region. }
\end{center}
\end{figure}
\begin{itemize}
\item Both of the surfaces at $r^{\ast}=\epsilon_{0}$ are spacelike. Using \eqref{GHYspace} we find
\begin{align}
I_{GHY}^{XW} &=   \frac{-r^{d-1} ~\Omega_{d-1}}{16 \pi G_{n}} ~\left( f'(r) +\frac{2(d-1)}{r} f(r) \right) ~\left(t(W)-t(X)\right),
\\
I_{GHY}^{YZ} &=   \frac{-r^{d-1} ~\Omega_{d-1}}{16 \pi G_{n}} ~\left( f'(r) +\frac{2(d-1)}{r} f(r) \right) ~\left(t(Z)-t(Y)\right).
\end{align}
Summing these two terms results in the contribution of $r^{\ast} = r^{\ast}(\epsilon_{0})$ surfaces,
\begin{equation}
I_{GHY}^{sing}= \frac{r^{d-1} ~\Omega_{d-1}}{16 \pi G_{n}} ~\left( f'(r) +\frac{2(d-1)}{r} f(r)\right) ~ (4 r^{\ast}(\epsilon_{0})+ 2 \mu^+-2\mu^-).
\end{equation}
 Using \eqref{schwmet} and \eqref{symmetricquantity}  we have
\begin{equation}
\frac{d I_\text{GHY}^\text{sing}}{d \mu}=0.
\end{equation}
Thus, in the symmetric case, the GHY terms of the near  singularities cancel each other and are independent of the boundary time.
\item Null joint terms take place at $P$ and $Q$,
\begin{align}\label{pointP}
I_{J}^{P} &= \frac{-\Omega_{d-1}~ r_P^{d-1}}{8 \pi G_{N}} ~\log \frac{\mathfrak{a}^2}{|f(r_P)|},
\\
\label{pointQ}
I_{J}^{Q} &= \frac{-\Omega_{d-1} ~r_Q^{d-1}}{8 \pi G_{N}} ~\log \frac{\mathfrak{a}^2}{|f(r_Q)|}.
\end{align}
Using $r_P^{\ast}=r_Q^{\ast}=0$ and \eqref{tort} we have $\frac{d~ r_P}{d \mu}=\frac{d~ r_Q}{d \mu}=0$. Therefore the time derivative of null-joint terms at these points vanishes
\begin{equation}
\frac{d I_{J}^\text{null}}{d \mu}=0.
\end{equation}

\item There are four spacelike joint terms. All of these joint terms take place near the singularities and are independent of the boundary time
\begin{align}
I_{J}^{X}&= \frac{\Omega_{d-1}~ r^{\ast}(\epsilon)^{d-1}}{16 \pi G_{N}} ~\log \big| f(r^{\ast}(\epsilon))\big|,
\\
I_{J}^{W}&=I_{J}^{Y}=I_{J}^{Z}=I_{J}^{X}.
\end{align}
\end{itemize}
 
 Putting all together we find 
\begin{equation}
\frac{d I}{d \mu}=0 \Rightarrow \dot{\mathcal{C}}=0.
\end{equation}
Hence the rate of  complexity growth  for the initial time is zero.
\subsection{Late time}

For the late times that are after critical time \eqref{criticaltime}, the Penrose diagram is depicted in figure 2. Similar to the initial time we want to calculate on-shell action in a region of spacetime that is determined by the gray area  in figure 2. This region is still the intersection of four WDW patches with UV and IR cutoffs. The distinct difference with the initial time is that 
 the cutoff surface at the past singularity does not exist,  and  we have to consider the null joint term in  the action at a new point $M$:
\begin{figure}\label{fig:a1}
\begin{center}
\includegraphics[scale=0.4]{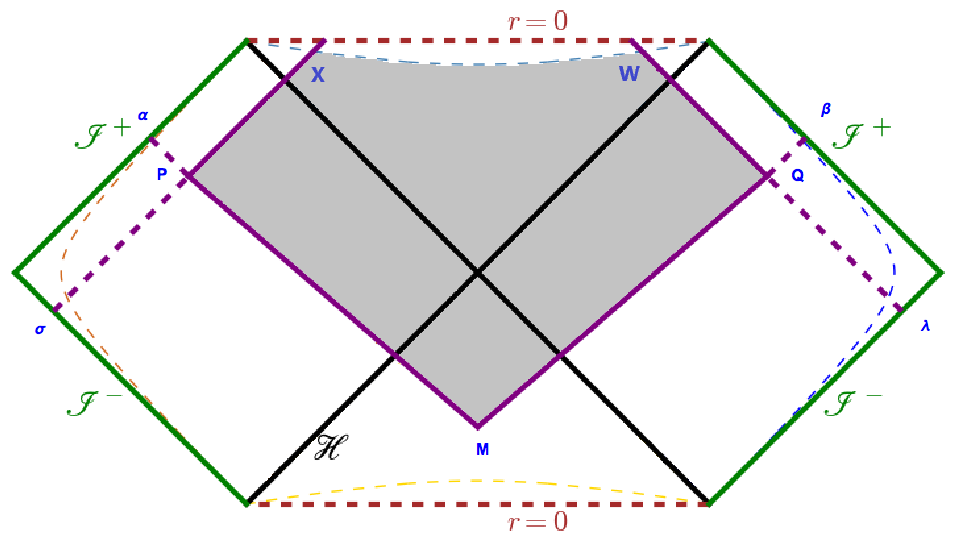}
\caption{ Late time  for $d>3$. Gravitational action is evaluated on the gray region.}
\end{center}
\end{figure}
\begin{itemize}

\item The only GHY term that contributes to our problem takes place near the future singularity
\begin{equation}\label{GHYlate}
I_{GHY}^{sing}=\frac{\Omega_{d-1} ~d ~\omega^{d-2}}{16 \pi G_{N}}~ (\mu - \chi)
\end{equation}
\item The contribution  of points $P$ and $Q$ is similar to the initial time, and it is vanishing.

\item  For the null joint term at point $M$  we have
\begin{equation}
\label{jointM}
I_{J}^{null}(M) =  \frac{r_M^{d-1} ~\Omega_{d-1}}{8 \pi G_{N}} \log  \left| \frac{\mathfrak{a}^2}{f(r_M)} \right|
\end{equation}
It is assumed that $t(M)=0$ at this point, which yields $r ^{\ast}(M)=-\mu^+ = -\frac{\mu+\chi}{2}$. The sign of $f(r_M)$ is negative, and  using  \eqref{tort}  it is not hard to find
\begin{equation}\label{mtime}
\frac{d r_M}{d \mu} = -\frac{ f(r_M)}{2}.
\end{equation}
 Using the previous equation we find
\begin{equation}
\frac{d I_{J}^\text{null}(M)}{d\mu} = \frac{r_M^{d-2} ~\Omega_{d-1}  }{16 \pi G_{N}} \left( r_M ~f'(r_M)-(d-1)f(r_M) \log \frac{-\mathfrak{a}^2}{f(r_M)}\right).
\end{equation}
  Using  \eqref{schwmet} we have
\begin{align}\label{pointM}
\frac{d I_{J}^\text{null}(M)}{d\mu} = \frac{(d-2) ~\omega^{d-2}~ \Omega_{d-1}}{16 \pi G_{N}}-\frac{r_M^{d-2} ~\Omega_{d-1} }{8 \pi G_{N}} \left((d-1)~f(r_M) ~\log \frac{-\mathfrak{a}^2}{f(r_M)}\right).
\end{align}
Using \eqref{GHYlate}, the  boundary contribution is given  by  
\begin{equation}\label{lateboundary}
\frac{d I_\text{GHY}^\text{sing}(M)}{d\mu} =\frac{\Omega_{d-1} ~d ~\omega^{d-2}}{16 \pi G_{N}}.
\end{equation}
Finally, the rate of  complexity growth in the  flat  case  can be found by adding the last two terms \eqref{pointM} and \eqref{lateboundary},
\begin{equation}\label{Scwlate}
\dot{\mathcal{C}} = \frac{1}{\pi}\frac{d I}{d \mu}=\frac{1}{\pi}\left[ 2 M - \frac{r_M^{d-2} ~\Omega_{d-1} (d-1)}{16 \pi G_{N}  }~f(r_M)\log \frac{-\mathfrak{a}^2}{f(r_M)}\right]
\end{equation} 
This is exactly \eqref{result for d4} which is given by taking the flat-space limit. 
\end{itemize}
In the symmetric evolution $r_M$ is always less than the horizon radius and   $-\infty<f(r_M)<0$. Using this fact, we find that  $\dot{\mathcal{C}}$  given by \eqref{Scwlate} starts from $-\infty$ and increases to a maximum value that is greater than $2M/\pi$ but finally at late times approaches  $2M/\pi$. In this view, the Lloyd's  bound  \cite{Lloyd:limit, Margolus:1997ih} is approached from above for all values of parameter $\mathfrak{a}$. 

 \subsubsection{Numerical results for the Schwarzschild black hole}
In this subsection we  present the  numerical analysis of the complexity growth for the four-dimensional Schwarzschild metric.

The critical  time in this case can be found through the relations \eqref{schwmet}, \eqref{tort} and \eqref{criticaltime},
\begin{equation}
\mu_c = -\omega \log( \omega ).
\end{equation} 
$r_M$ can be calculated using \eqref{tort} where for $d=3$ we have
\begin{equation}
r_{M} + \omega ~\log ( r_{M} - \omega ) = -\frac{\mu+\chi }{2}.
\end{equation}
Using this equation we  find $r_M$ as
\begin{equation}
r_{M} = \omega ~\left( 1 + W\left( \frac{-1}{\omega ~e^{1+ \frac{\mu + \chi}{2\omega}}}\right ) \right),
\end{equation}
where $W$ is the Lambert $W$ function.
The rate of complexity growth at late time can be read from \eqref{Scwlate} by putting $d=3$,
\begin{equation}
\dot{\mathcal{C}} =\frac{1}{\pi}\left( 2 M - \frac{r_{M}}{2 G_{N}} f(r_M) \log \frac{-\mathfrak{a}^2}{f(r_M)}\right).
\end{equation}
For a fixed value of horizon parameter $\omega$ but different values of $\mathfrak{a}$ the complexity growth with respect to the boundary time is plotted in figure  3. For all  values of the parameters,  the rate of the complexity growth at late time approaches  Lloyd's bound \cite{Lloyd:limit, Margolus:1997ih}
from above.

\begin{figure}
\begin{center}
\includegraphics[scale=0.7]{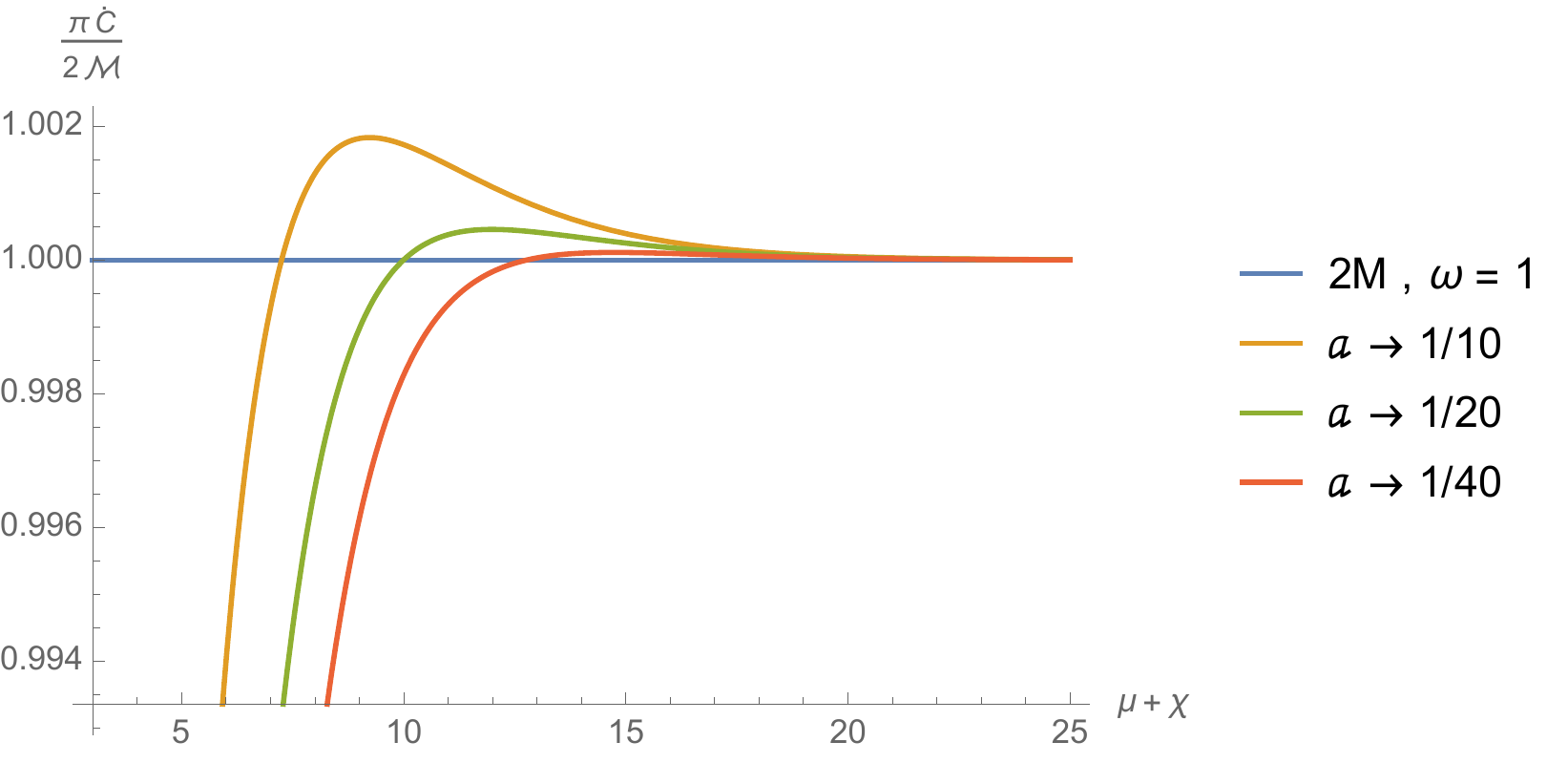}
\caption{The rate of complexity growth for different $\mathfrak{a}$.}
\end{center}
\end{figure}

\section{ BMSFT Complexity Growth in $d=2$}
 Asymptotically flat black hole solutions, which were studied in the previous section, are given by taking the  flat-space limit from the asymptotically AdS black holes. In the three-dimensional Einstein gravity, there is no black hole solution \cite{Ida:2000jh}. The flat space limit of BTZ black holes  are three-dimensional cosmological spacetimes \cite{Cornalba:2002fi,Cornalba:2003kd}. These solutions, which are known as flat space cosmology (FSC), have recently been   studied in the context of flat-space holography \cite{Bagchi:2012xr,Barnich:2012xq}.
In this section,   we  find the exact formula for the  growth rate  of the BMSFT complexity in $d=2$ \footnote{As mentioned in section 1, for $d>2$ the flat-space limit of the complexity growth rate has already been discussed in the context of the CV conjecture \cite{Susskind:2014moa}. Consequently, this section is the most original part of the paper where the details of the calculations are quite different from those of the AdS case.  }. To do so, we apply the CA conjecture for the  nonrotating   FSC given by 
\begin{align}\label{FSCmetric}
ds^2 &= \hat{r}_{+}^2 d t^2 - \frac{d r^2}{\hat{r}_{+}^2} +r^2 d \phi^2,
\\
M&=\frac{\hat{r}_{+}^2}{8 G_{N}},
\end{align} 
where $M$ is the mass parameter.
For this metric the  tortoise coordinate is given by 
\begin{equation}\label{tortoisefsc}
r^{\ast} = - \frac{r}{\hat{r}_{+}^2}.
\end{equation}
The Penrose-Cartan diagram of FSC has been shown in figure 4.
We note that all of the  $r=\text{const}$ surfaces are spacelike in  FSC.
Similar to the higher dimensional cases,   we  impose a symmetric dynamic  for  the right- and  left-hand sides of the Penrose-Cartan diagram. The necessary information for   all of the   relevant points in FSC are collected in the following table:
\begin{center}
\begin{tabular}{ |p{3cm}|p{3cm}|p{3cm}|p{2cm}|p{3.7cm}|  }
 \hline
 \multicolumn{4}{|c|}{The points on Penrose-Cartan diagram of FSC space} \\
 \hline
\quad Point name & \quad\qquad $t$ &\quad\qquad $r ^{\ast}$ & Sign of $f(r)$  \\
 \hline
\quad\qquad X &\quad $r^{\ast}(\epsilon_{0}) - \mu^-$ & \qquad$r^{\ast}(\epsilon_{0})$  &\qquad -\\
 \hline
 \quad\qquad W &\quad $-r^{\ast}(\epsilon_{0}) + \mu^-$ &\qquad $r^{\ast}(\epsilon_{0})$  &\qquad - \\
 \hline
\quad\qquad Y &\quad $r^{\ast}(\epsilon_{0}) + \mu^+$ &\qquad $r^{\ast}(\epsilon_{0})$  &\qquad - \\
 \hline
\quad\qquad Z &\quad $-r^{\ast}(\epsilon_{0}) - \mu^{+}$ &\qquad $r^{\ast}(\epsilon_{0})$  &\qquad -  \\
  \hline
\end{tabular}
\end{center}
\vspace{.5cm}
The bulk action and also the boundary terms and the joint terms are exactly the same as in \eqref{action}. The definition of the initial and the late times are also the same as for higher dimensional cases. 
 \subsection{Initial time}
 \begin{figure}
\label{fig:22}
\begin{center}
\includegraphics[scale=0.4]{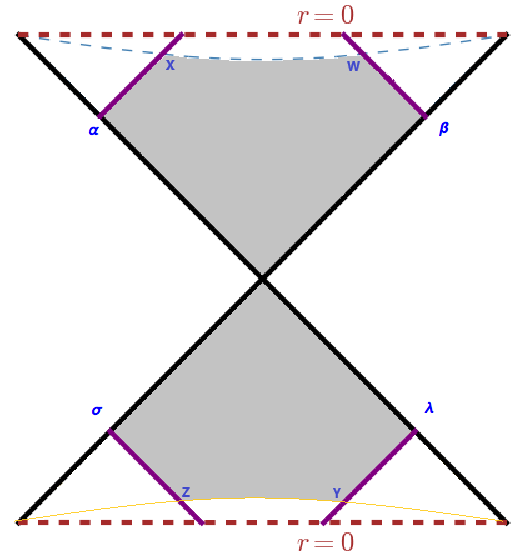}
\caption{Initial time for FSC. Gravitational action is evaluated on the gray region.}
\end{center}
\end{figure}
 The GHY term for the spacelike surfaces in the FSC \eqref{FSCmetric} are
 \begin{equation}
I_{GHY}^{FSC}= \int \frac{\hat{r}_{+}^2}{4 G_{N}} dt \Bigg{|}_{r^{\ast} = \text{const}}.
 \end{equation}
For the surfaces near  $r^{\ast} = \epsilon_{0}$ we have 
\begin{align}
I_{GHY}(WX) &= \frac{\hat{r}_{+}^2}{4 G_{N}} \left( t(W)- t(X)\right ),
\\ 
I_{GHY}(ZY) &= \frac{\hat{r}_{+}^2}{4 G_{N}} \left( t(Z)- t(Y)\right ).
\end{align}
Adding these two terms yields
\begin{equation}
I_{GHY}^{sing} = \frac{\hat{r}_{+}^2}{4 G_{N}}\left ( -4 r^{\ast}(\epsilon_0) + 2 \mu^- -2 \mu^+ \right).
\end{equation}
Hence for the symmetric evolution where  $\mu^+ - \mu^-$ is fixed,  the rate of complexity growth for the initial time vanishes,
\begin{equation}
\dot{\mathcal{C}}_{initial}^{FSC} = 0.
\end{equation}
\subsection{Late time}
\begin{figure}\label{fig:22}
\begin{center}
\includegraphics[scale=0.4]{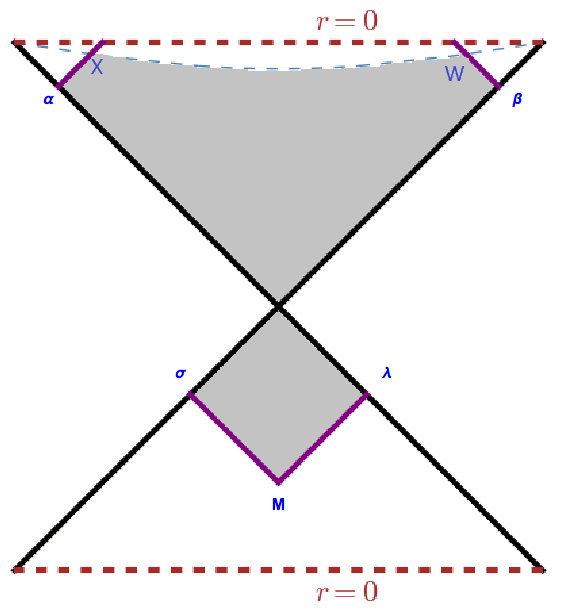}
\caption{Late time for FSC. Gravitational action is evaluated on the gray region.}
\end{center}
\end{figure}

At the late time a new  null joint term appears  at the point $M$ (Figure 5). At this point $t(M)=0$, $r ^{\ast}(M)=- \mu^+$, and the  sign of $f(r_M)$ is negative. Thus we have
\begin{equation}
\frac{d I_{J}(M)}{d \mu^+} = -\frac{\hat{r}_{+}^2}{8 G} \log |\frac{\hat{r}_{+}}{\mathfrak{a}}|.
\end{equation}
 Imposing the symmetric evolution,
\begin{equation}
\mu^+ - \mu^- = \text{const}, \qquad \mu^+ + \mu^-={\mu},
\end{equation}
the complexity growth  at the late time is found, 
\begin{equation}\label{final FSC}
\dot{\mathcal{C}}=\frac{1}{\pi}\frac{d I}{d \mu} = \frac{1}{\pi} \left( \frac{\hat{r}_{+}^2}{4 G_{N}} + \frac{\hat{r}_{+}^2}{4 G_{N}} \log |\frac{\mathfrak{a}}{\hat{r}_{+}}|\right).
\end{equation}
We can rewrite  \eqref{final FSC} in terms of   the FSC mass as
\begin{equation}
\dot{\mathcal{C}}= \frac{1}{\pi} \left( 2 M+ M \log \frac{\mathfrak{a}^2}{8 G_{N} M}\right).
\end{equation}
It is a constant and differs from  Lloyd's bound by a logarithmic term. This result is exactly the same as \eqref{result d3}, which is given by taking the flat space limit. Similar to the higher dimensional cases, $\mathfrak{a}$ is a new length parameter in the field theory. However, if we demand that  Lloyd's bound \cite{Lloyd:limit} is not violated, then $\mathfrak{a}$ must be restricted by   $\mathfrak{a}< \hat{r}^{+}$. 
\section{Discussion } 
In this paper we calculate the rate of complexity growth for BMSFT$_d$. Our main goal is generalizing the CA proposal for the  flat/BMSFT correspondence. To do so, we need to define a portion of spacetime in which the gravitational action is evaluated. Since the final formulas for the growth rate  are simply given  by taking the flat space limit from the AdS/CFT calculation, we can check the results of various potential regions. Our main achievement in this paper is that the evaluation of the gravitational action must be done on a region that is given by the intersection of four WDW patches. The boundary of two of these patches connects past singularity to the future one by crossing from  past null infinity. The boundary of other patches starts from past null infinity and using future null infinity eventually reaches to the future singularity. Using this portion of spacetime we show that the rate of complexity growth is zero for the initial times and for the late times  is exactly the same as what one finds by taking the flat-space limit from the AdS/CFT calculation.

Despite  CFT$_2$, the rate of complexity growth is a constant for the BMSFT$_2$ and differs from  Lloyd's bound \cite{Lloyd:limit} by a logarithmic term. In higher dimensions, this rate is not a constant and approaches  Lloyd's bound from above.   This result is similar to  the  higher dimensional asymptotically AdS black holes  and different  from the Lifshitz and the hyperscaling violating geometries that this bound has violated \cite{Alishahiha:2018tep}. This shows that although BMSFTs are ultrarelativistic theories, they are more similar to the relativistic theories than to the nonrelativistic ones.

 The idea presented in this paper for the  definition of the suggested  spacetime region on which the gravitational  action is evaluated may help us to study the complexity of formation \cite{Brown:2015bva, Brown:2015lvg, Chapman:2016hwi} in  the context of  flat/BMSFT correspondence. This is a potentional open problem that could be addressed in  future works.
 
\section*{Acknowledgements}
The authors  thank S.~M.~Hosseini and Sh. Karimi for useful comments on the manuscript. We are also grateful to M.~R.~Mohammadi Mozaffar, A.~Faraji Astaneh and Yang Run Qiu for the useful comments.
 We have used xAct \cite{xAct} and its package\cite{Nutma:2013zea} during this work and for Penrose-Cartan diagrams we have used the code by  E. Gourgoulhon \cite{sage}. 
 
 \appendix
\section{ Non-symmetric evolutions in past and future null infinities}
In this appendix, we discuss the nonsymmetric scenario when the boundary times on future and past null infinities develop individually. We show  the parameter that  render nonsymmetricity by $\gamma$. Hence we define
\begin{equation}
\mu^+ = \gamma \mu^- + \chi  \qquad \mu^+ + \mu^- =\mu,
\end{equation}
 where $\chi$ is a constant. Using nonsymmetric parametrization, the rate of  complexity growth is  given by
\begin{equation}\label{deriv}
\dot{\mathcal{C}}=\frac{1}{\pi}\left(\frac{d I}{d \mu}\right) = \frac{1}{\pi}\left(\frac{\gamma}{1+\gamma}\frac{d I}{d \mu^+}  + \frac{1}{1+\gamma}\frac{d I}{d \mu^-}\right).
\end{equation}
\subsection*{A1 Initial time}
In this case the  GHY terms near the  singularities change as 
\begin{equation}
I_{GHY}^{sing}=\frac{\Omega_{d-1} ~d~ \omega^{d-2}}{8 \pi G_{N}}\left( \mu^{-} - \mu^+\right).
\end{equation}
Using the fact that $r_P=r_Q$ and also  \eqref{pointP} and \eqref{pointQ}
the null joint terms become
\begin{equation}
I_{J}^{null} = -\frac{\Omega_{d-1} ~(r_P )^{d-1}}{4 \pi G_{N}} \log \frac{\mathfrak{a}^2}{|f(r_P)|}.
\end{equation}
Using the  definition of the tortoise coordinate \eqref{tort}, one can find
\begin{equation}
\frac{d r_P}{d \mu^+}= \frac{-f(r_P)}{2}, \qquad \frac{d r_P}{d \mu^-}= \frac{f(r_P)}{2}.
\end{equation}
Hence we find the rate of complexity growth for nonsymmetric boundary times as
\begin{align}
\dot{\mathcal{C}} =\frac{1}{\pi} ~\frac{d I}{d \mu}=\frac{1-\gamma}{1+\gamma} ~\frac{1}{\pi} ~ \left(4 M+ \frac{\Omega_{d-1} ~(d-1)~r_P^{d-2}}{8 \pi G_{N}} f(r_P) ~\log \frac{\mathfrak{a}^2}{f(r_P)}\right).
\end{align}
It is clear that for the symmetric case, i.e.,   $\gamma=1$, the growth rate is zero.
\subsection*{A2 Late time}
Now the boundary term near the  future singularity results in 
\begin{equation}\label{GHYlateap}
I_{GHY}^{sing}=\frac{\Omega_{d-1} ~d~ \omega^{d-2}}{8 \pi G_{N}}( \mu^-).
\end{equation}
The contribution of the joint terms at $P$ and $Q$ is similar to the initial time where 
we have
\begin{equation}
I_{J}^{null} = -\frac{\Omega_{d-1} ~r_P ^{d-1}}{4 \pi G_{N}} \log \frac{\mathfrak{a}^2}{|f(r_P)|}.
\end{equation}
The contribution of the null-joint term at point $M$ is given by \eqref{jointM}. Using \eqref{tort} it is not hard to find that for the nonsymmetric case we have
\begin{equation}
\frac{d r_M}{d \mu^+} = -f(r_M).
\end{equation}
Hence we find 
\begin{align}
\frac{d I_{J}^\text{null}(M)}{d \mu^+} = \frac{(d-2) ~\omega^{d-2} ~\Omega_{d-1}}{8 \pi G_{N}}-\frac{r_M^{d-2}~\Omega_{d-1}}{8 \pi G_{N}} ~\left((d-1)f(r_M) \log \frac{-\mathfrak{a}^2}{f(r_M)}\right).
\end{align}
Finally  using \eqref{deriv} we find the rate of complexity growth as
\begin{align}
\nonumber\dot{\mathcal{C}} =  \frac{1}{\pi} ~\Big[ \frac{4 M}{1+\gamma}  &- \frac{r_M^{d-2}~ \Omega_{d-1}}{8 \pi G_{N}} \frac{\gamma}{1+\gamma}~(d-1)f(r_M) \log \frac{-\mathfrak{a}^2}{f(r_M)} 
\\
&+\frac{1-\gamma}{1+\gamma} \frac{\Omega_{d-1} ~(d-1)~r_P^{d-2}}{8 \pi G_{N}} f(r_P) \log \frac{\mathfrak{a}	^2}{f(r_P)} \Big].
\end{align}
For $\gamma=1$ the result is exactly the same as the symmetric case. Moreover, in the nonsymmetric case, the complexity growth depends not only on  $r_M$ but also on  $r_P$.

\end{document}